\documentclass[10pt,aps,pra,groupedaddress,showpacs,twocolumn,nofootinbib,superscriptaddress,longbibliography,amsmath,amssymb,floatfix]{revtex4-1}
\usepackage{xcolor}
\usepackage{graphicx}
\usepackage{float}
\usepackage{verbatim}
\usepackage[colorlinks=true, linkcolor=blue, urlcolor=blue, citecolor=blue, hyperindex, breaklinks]{hyperref}
\usepackage[normalem]{ulem}
\usepackage{algpseudocode}
\usepackage{algorithmicx}
\usepackage{algorithm}
\usepackage{microtype}
\usepackage[capitalise]{cleveref}

\usepackage{pgf}

\newcommand{\distance}{}
\newcommand{\distanceroman}{}

\newcommand\add[1]{#1}

\newcommand{\setthreshold}[4]{
  \expandafter\pgfmathsetmacro\csname hdrg#1\endcsname{#2}
  \expandafter\pgfmathsetmacro\csname mwpm#1\endcsname{#3}
  \expandafter\pgfmathsetmacro\csname ensemble#1\endcsname{#4}
}

\setthreshold{V}{6.54}{8.05}{11.14}
\setthreshold{VII}{6.85}{10.65}{12.21}
\setthreshold{IX}{8.63}{12.01}{12.86}
\setthreshold{XI}{9.22}{12.59}{13.07}

\newcommand{\printthreshold}[1]{%
\pgfmathprintnumber[fixed, precision=2, fixed zerofill=true]{#1}}

\newcommand{\printpercent}[3][2]{%
  \pgfmathsubtract{#2}{#3}%
  \pgfmathdivide{\pgfmathresult}{#3}%
  \pgfmathmultiply{\pgfmathresult}{100}%
  \pgfmathprintnumber[fixed, precision=#1, fixed zerofill=true]{\pgfmathresult}\%}

\newcommand{\set}[1]{\left\{ #1 \right\}} 

\newcommand{\cQ}{\mathcal{Q}}
\newcommand{\Z}{\mathbb{Z}}

\begin{document}

\title{Neural ensemble decoding for topological quantum error-correcting codes}

\author{Milap Sheth}
\email[Electronic address: ]{m2sheth@uwaterloo.ca}
\affiliation{Institute for Quantum Computing, University of Waterloo, Waterloo, ON, Canada}
\affiliation{Department of Combinatorics and Optimization, University of Waterloo, Waterloo, ON, Canada}
\affiliation{David R. Cheriton School of Computer Science, University of Waterloo, Waterloo, ON, Canada}

\author{Sara Zafar Jafarzadeh}
\email[Electronic address: ]{zadeh.sarah@gmail.com}
\affiliation{Institute for Quantum Computing, University of Waterloo, Waterloo, ON, Canada}
\affiliation{Department of Computer Science and Operations Research, Universit\'e de Montreal, Montreal, QC, Canada}

\author{Vlad Gheorghiu}
\email[Electronic address: ]{vlad.gheorghiu@uwaterloo.ca}
\affiliation{Institute for Quantum Computing, University of Waterloo, Waterloo, ON, Canada}
\affiliation{Department of Combinatorics and Optimization, University of Waterloo, Waterloo, ON, Canada}
\affiliation{softwareQ Inc., Kitchener, ON, Canada}


\begin{abstract}
  Topological quantum error-correcting codes are a promising candidate for building fault-tolerant quantum computers.
  Decoding topological codes optimally, however, is known to be a computationally hard problem. Various decoders have been proposed that achieve approximately optimal error thresholds. Due to practical constraints, it is not known if there exists an obvious choice for a decoder.
  In this paper, we introduce a framework which can combine \emph{arbitrary} decoders for any given code to significantly reduce the logical error rates.
  We rely on the crucial observation that two different decoding techniques, while possibly having similar logical error rates, can perform differently on the \emph{same} error syndrome.
  \add{We classify each error syndrome to the decoder which is more likely to decode it correctly using machine learning techniques.}
  We apply our framework to an ensemble of Minimum-Weight Perfect Matching (MWPM) and Hard-Decision Re-normalization Group (HDRG) decoders for the surface code in the depolarizing noise model.
  Our simulations show an improvement of \printpercent[1]{\ensembleV}{\mwpmV}, \printpercent[1]{\ensembleVII}{\mwpmVII}, and \printpercent[1]{\ensembleIX}{\mwpmIX} over the pseudo-threshold of MWPM in the instance of distance 5, 7, and 9 codes, respectively.
  Lastly, we discuss the advantages and limitations of our framework and applicability to other error-correcting codes.
  Our framework can provide a significant boost to error correction by combining the strengths of various decoders. In particular, it may allow for combining very fast decoders with moderate error-correcting capability to create a very fast ensemble decoder with high error-correcting capability.
\end{abstract}

\maketitle

\section{Introduction}
\label{sct:intro}
Quantum computing is a disruptive new model of computation that relies on the core principles of quantum mechanics for performing specific computational tasks, some significantly (sometimes exponentially) faster than any known classical algorithm. Some prominent examples of algorithms that achieve ``quantum speedups'' are Shor's algorithm for factoring~\cite{SJC.26.1484}, which achieves an exponential speedup compared to the best classical algorithm based on number-sieving,
or Grover's algorithm for searching un-ordered spaces~\cite{PhysRevLett.79.325}, which achieves a quadratic speedup compared to brute force searching, while being optimal~\cite{PhysRevA.60.2746}.

The most significant obstacle for practical realization of quantum
computers is the high level of noise (or decoherence) that affects
the individual physical qubits required by the computation. Initially considered as an insurmountable problem due to the no-cloning theorem, it was subsequently proven that quantum error correcting codes (QECC)
exist~\cite{PhysRevA.54.1098}. In QECC, a large number of
physical qubits are encoded into a logical qubit, the latter
having a smaller error rate than the former. Moreover, due to the threshold
theorem~\cite{Aharonov:1997:FQC:258533.258579}, the effective logical error
rate can be made arbitrarily small, which enables arbitrary
quantum computation at the price of increasing the number of physical qubits.

Surface codes~\cite{fowler_surface_2012}, which are an instance of topological quantum error correcting codes, the latter introduced by Kitaev~\cite{Kitaev20032}, are currently among the leading candidates for large scale quantum error correction. They are an instance of the more general class of stabilizer quantum error-correcting codes~\cite{quantph.9705052}.

However, optimal decoding for topological codes
is known to be a computationally hard problem~\cite{PhysRevA.74.052333}. Various decoders~\cite{PhysRevLett.104.050504, wootton_high_2012, torlai_neural_2017, delfosse2017almostlinear,PhysRevE.97.051302} have been proposed that
achieve approximately optimal error thresholds instead.
Due to several constraints (e.g. error-correcting capability, low coherence time, near absolute-zero temperature hardware environment etc.), each decoder suffers some disadvantage.
Hence, it is not clear which decoder is the best candidate in practice.

In this paper, we introduce a framework for combining \emph{arbitrary} decoders for any specific topological code
to significantly reduce the error rates. We rely on the crucial observation that two different
decoding techniques, while possibly having similar error rates, can perform differently on the \emph{same} error syndrome.
We use classification techniques from machine learning to show that an ensemble of decoders can correct more errors
than any decoder individually. We apply our techniques to the Minimum-Weight Perfect Matching (MWPM)~\cite{edmonds_paths_1965, dennis_topological_2002} and Hard-Decision Re-normalization Group (HDRG)~\cite{PhysRevLett.104.050504} decoders for the surface code
in the depolarizing noise model.
As summarised in \cref{tab:thres}, our simulations show an improvement of \printpercent[1]{\ensembleV}{\mwpmV}, \printpercent[1]{\ensembleVII}{\mwpmVII}, \printpercent[1]{\ensembleIX}{\mwpmIX}, and \printpercent[1]{\ensembleXI}{\mwpmXI} over the pseudo-threshold of MWPM in the instance of distance 5, 7, 9, and 11 codes respectively.

Lastly, we discuss the advantages and limitations of our approach and possible avenues for improvements.
Our framework can provide a significant boost to error correction by combining decoders that exploit different characteristics of the error syndrome.
In particular, it may allow for combining very fast decoders with moderate error-correcting capability to create a very fast ensemble decoder with high error-correcting capability.

The remainder of this paper is organized as follows.
In \cref{sct:preliminaries} we give an overview of the decoding problem on the surface code.
In \cref{sct:methods} we describe our framework for creating an ensemble of decoders using machine learning techniques.
In \cref{sct:results} we discuss our results based on using an ensemble of decoders for error correction.
Finally, \cref{sct:conclusion} concludes our paper and raises a series of open questions.

\section{Preliminaries}
\label{sct:preliminaries}

\begin{figure}
  \centering
  \includegraphics[scale=0.8]{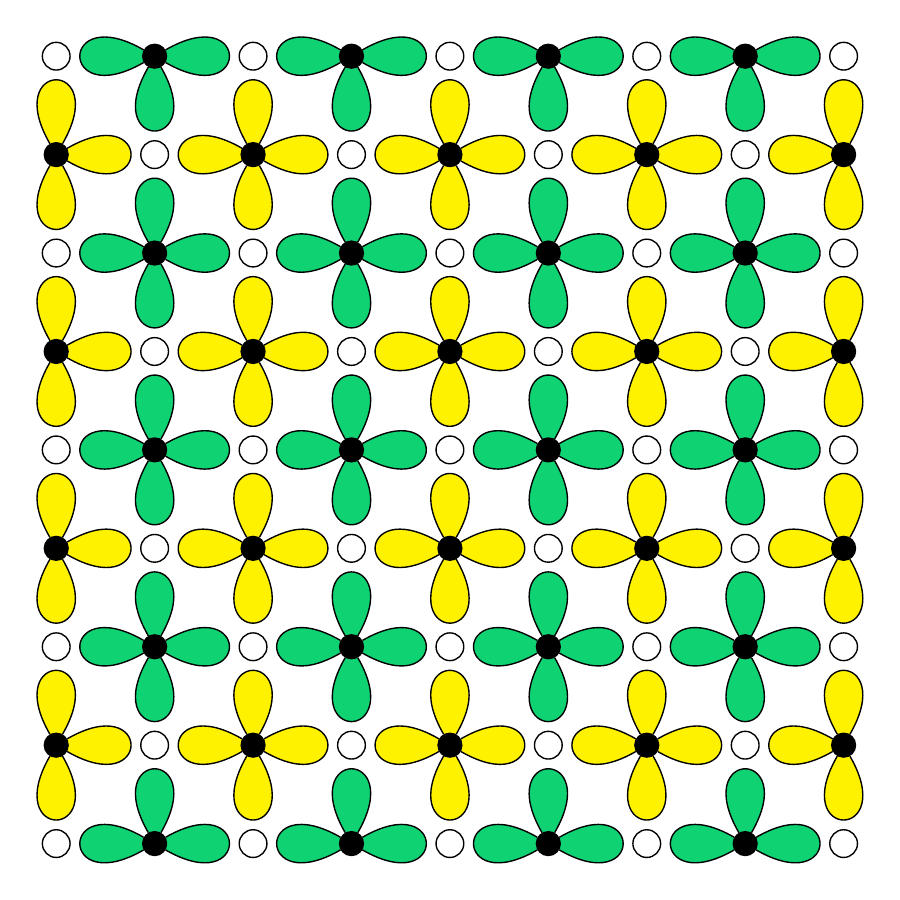}
  \caption{Layout of the distance 5 surface code. White circles are the data qubits and black circles are the measurement qubits. $Z$ and $X$ stabilizers are indicated by the green and yellow petals, respectively. Each stabilizer is coupled with its adjacent data qubits.}
  \label{fig:surface}
\end{figure}

The \emph{surface code} is a family of topological stabilizer codes having data qubits
arranged in a 2-dimensional lattice at the edges and measurement qubits lying on the vertices and faces.
A distance $ L $ surface code uses $ n = L^2 + (L-1)^2 $ data qubits and
$ m = 2 L (L - 1) $ measurement qubits for a total of $ N = (2 L - 1)^2 $ qubits to encode a single logical qubit.
As shown in \cref{fig:surface}, the qubits corresponding to the yellow petals, labelled by $ v $, represent the $ X $-stabilizers and the qubits corresponding to the green petals, labelled by $ f $, represent the $ Z $-stabilizers of the data qubits, respectively.
Thus, each data qubit not at the boundary is coupled to two $ X $ and two $ Z $ stabilizers.
An $ X $-stabilizer detects a Pauli $ Z $ error on an adjacent data qubit by showing a flipped measurement result (compared to its value at the previous time step).

Let $ \cQ $ denote the set of data qubits and $ \delta(p) $ denote the set
of data qubits coupled with a vertex or a face labelled by $ p $.
A Pauli operator on data qubit $ q \in \cQ $ in the surface code is denoted by $ P_q \in \set{I_q, X_q, Y_q, Z_q} $.
A Pauli error on the data qubits is then denoted by $ \prod_{q \in \cQ}{P_q} $, excluding the tensor product of identity operators.
The vertex and face stabilizer operators can be defined as
$ A_v = \otimes_{i \in \delta(v)}{X_i} $ and $ B_f = \otimes_{i \in \delta(f)}{Z_i} $, respectively.

Decoding of a surface code involves measuring the stabilizers on every surface code cycle and recording
changes in the corresponding eigenvalues obtained from the measurement results.
We define the \emph{error syndrome} of a surface code cycle to be the change in measured eigenvalues with respect to its state in the previous time step.
We denote the error syndrome by $ \vec{s} \in \Z_2^m $, where $ \vec{s}_i = 1 $ indicates a change in the eigenvalue of the $ i $th stabilizer.
We denote the error syndrome space $ \Z_2^m $ by $ \mathcal{S} $.

A logical Pauli $ X $ (similarly for $ Z $) error occurs when the Pauli $ X $ errors form a chain that
connects opposite sides of the boundary, as such a chain results in no apparent change of the corresponding measured error syndrome.
The decoding problem is hence defined as identifying the correct error operator after each code cycle
based on the error syndrome, see e.g.~\cite{fowler_surface_2012, dennis_topological_2002} for a comprehensive introduction.
The \emph{pseudo-threshold} corresponding to a specific distance code and decoder can then be defined as the
maximum error probability below which the logical error rate is lower than the physical error rate~\cite{svore_flow-map_2006}.
Fault-tolerant quantum computing is achieved below pseudo-threshold error rates as a result
of bootstrapping the error-correction mechanism by encoding logical qubits in another code.

\section{Ensemble decoding}
\label{sct:methods}

There have been various decoders proposed for the surface code, e.g.~\cite{PhysRevLett.104.050504, wootton_high_2012,torlai_neural_2017,delfosse2017almostlinear,PhysRevE.97.051302}.
Each decoder comes with its own set of advantages and disadvantages. Some of the important desideratum include but are not limited to error-correcting capability, decoding speed, or
ability to be implemented in hardware-specific environments (e.g.~in cryogenic environments for superconducting qubits)~\cite{fowler_surface_2012, chamberland_deep_2018}.
Finding the best decoder for a given set of constraints, specifically those ones that allow experimental realization, is an open problem.

In this paper, we improve upon the decoding performance of individual decoders by using an ensemble of decoders.
We begin by introducing some terminology.
Given a decoder $D$ and a list of error syndromes $S$, we define $S|_D$ to be the fraction of error syndromes in $S$ that $D$ decodes correctly.
Thus, a decoder's error-correcting capability corresponds to how large the fraction $S|_D$ is, where $S$ is chosen to be a large enough error syndrome list sampled based on the error model we are concerned with.

We observe that each decoder uses a subset of the error syndrome properties for decoding. This subset may vary from one decoder to the other.
The above observation implies that, given an error syndrome, while one decoder may fail to decode it, there might exist another decoder which correctly decodes it.

Let $S$ be a list of error syndromes and $\mathcal{D}$ be the set of decoders~$D_{\ell} \in \mathcal{D}$, for $ \ell \in \set{1, \hdots, k} $. Furthermore, let~$ E $ be a decoder that correctly decodes an error syndrome if and only if at least one of the decoders~$ D_{\ell} $ also corrects the same error syndrome. Hence, we obtain that,
\begin{equation}
\label{eqn1}
 S|_E \geq \max_{\ell}S|_{D_{\ell}}.
\end{equation}
In other words, given a set of decoders, the proportion of error syndromes correctly decoded by at least one decoder is greater than or equal to the proportion correctly decoded by the decoder with the highest error-correcting capability from the given set.

Therefore, we use the following method to construct an ensemble decoder.
Given a particular error syndrome, use the decoder which is the most likely to correctly decode it.
Hence, we have a classification problem where each error syndrome is labelled by the decoder that is the most likely to correct it.
Let~$\mathcal{D}$ denote the given set of~$k$ decoders and~$\mathcal{S}$ denote the error syndrome space~$ \Z_2^m $ as before.
The classification task is to learn the function~$ E: \mathcal{S} \to \mathcal{D} $,
where for any error syndrome $ \vec{s} \in \mathcal{S} $, $ E(\vec{s}) $ corresponds to the decoder that is the most likely to correctly decode it.

A naive solution to the above problem consists of performing a long enough Monte Carlo simulation of the error-correcting code where for each error syndrome the decoder that corrects it the highest number of times is recorded.
However, due to the exponential scaling of the  error syndrome space being, such an approach is computationally unfeasible for distances larger than 5.
To overcome this problem, we use neural networks to approximate the classification function discussed in the above paragraph. We assume a basic knowledge of deep learning applications to machine learning; for an introduction the reader can consult e.g.  Ref.~\cite{goodfellow_deep_2016}.
The input and output of our neural network is described below.

\begin{figure*}[t]
  \centering
  \includegraphics[width=\linewidth]{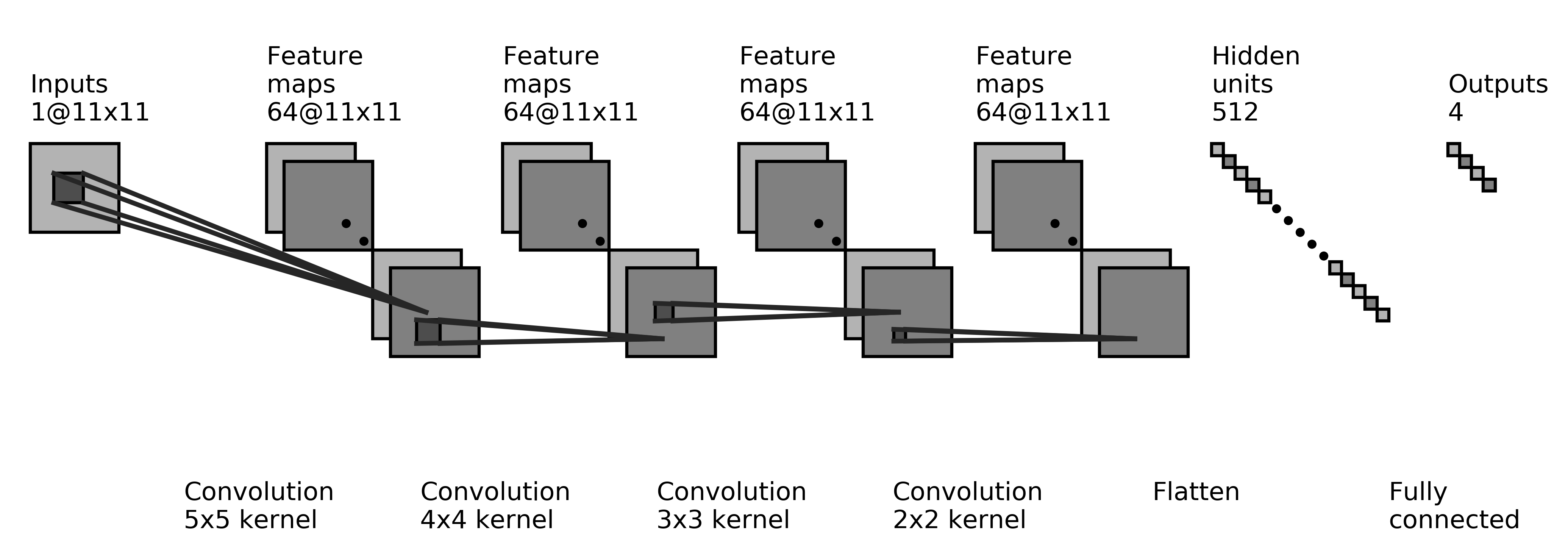}
  \caption{Neural Ensemble architecture for the distance 5 surface code.}
  \label{fig:cnn}
\end{figure*}

\emph{Input}. Error syndrome $ \vec{s} $ represented by $m$-dimensional binary vector
where a non-zero $ \vec{s}_i $ indicates a change in eigenvalue of the $ i $th $X/Z$ stabilizer.

\emph{Output}. Decoder $ D_{\ell} \in \mathcal{D} $ denoted by its label $ \ell \in \set{1, \hdots, k} $.

We first describe in \cref{sbsct:top} the architecture of a neural network that can be used for any topological code.
Then, in \cref{sbsct:surface}, we improve the architecture in the instance of the surface code by exploiting its lattice structure.
We then describe our training process and model evaluation
and some of our failed attempts for improving classification.
Subsequently, in \cref{sbsct:gen}, we briefly discuss generalizing our architecture to support other topological codes.
Finally, in \cref{sbsct:compare}, we compare our framework to existing neural decoding techniques.

\subsection{Topological codes}
\label{sbsct:top}

For a general topological code~\cite{Kitaev20032}, we use a deep fully-connected feed-forward neural network to learn the
labelling function mapping error syndromes to decoders most likely to correctly decode them.
Since a fully-connected feed-forward network does not assume any structural representation of the input error syndromes,
it can be applied to any topological code (and in fact any code assuming the existence of various decoders for it).

We successfully applied the above architecture to the distance 5 surface code during our preliminary investigations and obtained an increase of around 15\% over the pseudo-threshold of MWPM.
Our architecture consists of an input error syndrome layer,
4 hidden layers each with 512 neurons followed by a 256 neuron hidden layer,
and the final output layer with a neuron for each decoder.
We also utilized some of the optimizations mentioned in \cref{sbsct:train}.
Unfortunately, such an approach requires exceedingly large training data sets for surface codes with distances larger than 5 and quickly becomes unfeasible.

\subsection{Surface codes}
\label{sbsct:surface}

While the general architecture employed for arbitrary topological codes already provides significant results for the surface code,
we can improve it further by exploiting the structure of the latter.
We observe that, similarly to the task of image recognition, decoding an error syndrome of the surface code requires recognizing local features
and combining them to obtain global features.
This is precisely the strength of convolutional neural networks~\cite{726791}.
We utilize the translation invariance property of surface codes to feed the error syndromes to a convolutional neural network (CNN).
\add{Furthermore, the extent of local correlations in the error syndrome
  that we are concerned with depends on the distance of the surface code.
  This observation suggests that the size of the kernels in our CNN layers
  should be restricted by the distance of the surface code.}
The use of CNNs greatly reduces the number of trainable parameters and thus allows for more layers in the architecture.

CNNs operate on an image-like 2-dimensional input where neighbouring entries are correlated.
Conveniently, an error syndrome of the surface code already has a lattice structure where neighbouring error syndrome values are correlated.
In particular, the error syndrome has two lattices corresponding to the vertex and face stabilizers.
So we can map each error syndrome to a CNN input by rotating the error syndrome by $ 45 $ degrees,
obtaining a single lattice containing both type of stabilizers while preserving neighbouring correlations
and padding non-syndrome entries with zeros to obtain a matrix.
Note that this transformation is not necessary in the case of the rotated surface code as the error syndrome is already of the correct form,
with the exception of the boundary stabilizers for which zero padding is required.
Such a representation of the error syndrome also allows for detecting correlations between $ X $ and $ Z $ errors, improving the classification accuracy.

The transformed error syndrome is then fed through a sequence of convolutional layers of decreasing kernel sizes
(from $ L \times L $ to $ 2 \times 2 $) with 64 output filters each,
followed by one fully connected layer with 512 neurons, before producing the probability vector over the decoder labels.
\add{We use a sequence of decreasing kernel sizes to reduce the scope of local correlations corresponding to how errors are correlated in the surface code, and put more emphasis on the adjacent correlations in the final layer. This intuition is backed by our experiments of trying out different number of layers and combinations of kernel sizes. Such a decreasing sequence of layers performed the best for all distances in our experiments.}
This architecture is shown
in~\cref{fig:cnn}\footnote{This figure was generated by adapting the code from\\ \url{https://github.com/gwding/draw_convnet}}
as used in training the neural ensemble for the distance 5 surface code.
For larger code distances, the architecture is fine tuned in the number of parameters for each layer
but the core structure we employed remains the same.
We use \textsc{ReLU}~\cite{nair_rectified_2010} as the activation function for the hidden layers
and \textsc{softmax}~\cite{bridle_probabilistic_1990} for the output layer.
The \textsc{Adam} optimizer~\cite{kingma2014adam}, and particularly the \textsc{AMSGrad} variant~\cite{j.2018on}
allows for convergence of the learning rate in training large \add{data sets} and provides significant improvements.
Decoder labels are one-hot encoded and categorical cross-entropy loss is used for training.
\add{Although other state-of-the-art machine learning techniques can be used to obtain an architecture with higher performance, this design suffices to convey the power of our paradigm while avoiding further complexity.}

\subsection{Training process}
\label{sbsct:train}

The training samples are generated using a Monte Carlo simulation of the surface code.
The decoders are labelled according to preference of their usage.
For each simulated error syndrome, the results of decoding are recorded for all decoders.
In the event that all decoders either succeeded or failed,
we discard the corresponding error syndrome since any of the decoders could have been chosen without affecting the logical error rate.
This technique provides a significant boost to the classification, as the neural network is fed a smaller quantity of higher quality information,
leading to high accuracy and faster training times.
\add{We generated a data set containing 10 million of such filtered samples for training our neural network.}

We note that the chosen decoders could have differing
logical error rates which may lead to a bias in the training data towards one decoder. Therefore, we allow decoders to have different label weights.
This method proved useful in the case of an unfiltered \add{data set}. However,
in the case of the filtered samples described in the previous paragraph, the method does not provide a noticeable improvement.

To improve convergence, we reduce the learning rate interactively when decrements in the loss stagnate between epochs.
To avoid spending unnecessary time on large training sets, we stop training if the loss does not keep decrementing by a certain threshold in each epoch.
We observed that if a model is trained near the pseudo-threshold, it will perform maximally for all the considered noise rates.
Finally, given the trained ensemble $E$, decoding is performed according to \cref{algorithm}.

\begin{algorithm}[H]
    \caption{Neural ensemble decoder}
    \label{algorithm}
    \begin{algorithmic}[1]
        \Procedure{Decode}{$\vec{s}, E$} \Comment{Decode error syndrome $ \vec{s} $}
        \State $D_{\ell} \gets E(\vec{s})$ \Comment{Predicted decoder $ \ell $ for $ \vec{s} $}
        \State $ \mathcal{R} \gets D_{\ell}(\vec{s})$  \Comment{Recovery operator $ \mathcal{R} $}
        \State \textbf{return} $\mathcal{R}$
        \EndProcedure
    \end{algorithmic}
\end{algorithm}

We now discuss some approaches that we considered which did not provide any improvements.
Potentially, a neural network that overfits to the training data
can be improved by using Dropout layers~\cite{srivastava_dropout_2014}. Using Dropout layers with various
rates between convolutional layers did not improve the accuracy of the ensemble.
In image recognition problems, local changes in features represent similar information. However, in the case of error correction, a slight change in the error syndrome likely corresponds to a different error which might require the use of a different decoder.
For this reason, we did not use the max pooling layer.

We also tried clustering our training data for label generation.
The idea behind this method is that for each decoder, error syndromes likely to correspond
to that decoder can be further clustered into subgroups under some metric. Clustering the training data then allows for
creating additional sub-labels for each decoder, potentially fine-tuning the learning task. We tried this approach
with KModes clustering~\cite{huang_clustering_1997} used for categorical data to no avail.
An interesting open question is whether one might still be successful with this approach by using another metric instead of Hamming distance for evaluating closeness of two error syndromes.
We also note that since clustering is applied only to the training data, we can even use the information on the data qubits
as part of the metric and thus the clustering algorithm.

\subsection{Implementation}

Our code is implemented in Python using \textsc{Tensorflow}~\cite{abadi_tensorflow_2016} with
\textsc{Keras} as the front-end~\cite{chollet_keras_2015}. We build upon \textsc{QTop}~\cite{marks_comparison_2017},
a Python library for quantum error correction.
We use \textsc{Networkx}~\cite{aric_networkx_2008}, a Python library, for graph related functions.
We utilize the MWPM and HDRG decoders present in \textsc{QTop} for creating our ensemble.

\subsection{Generalizing the architecture}
\label{sbsct:gen}

The previous subsections indicate that one might be able to use a more sophisticated architecture for classifying the error syndromes of other topological codes using a neural ensemble.
The main restriction to using a similar convolution-based architecture is that the error syndrome of the code has to be representable in a 2-dimensional grid in a manner that encodes the local correlations of the possible stabilizer measurements for the surface code.
This appears to be feasible for the color code~\cite{bombin_topological_2006} where one can encode the
color information in separate channels of the input data.
Our architecture can even be extended to the 3-dimensional variants of the surface and color codes by the usage of 3-dimensional CNNs.

Another approach is to use a generalization of CNNs,
known as graph convolutional networks~\cite{kipf_semi_2017}, that directly operate on graphs rather than 2-dimensional grids.
This allows for encoding error syndromes of error-correcting codes with a more complicated topology.

\subsection{Comparison to neural decoders}
\label{sbsct:compare}

Finally, we briefly compare our framework to existing neural decoders for topological codes.
One of the common approaches that neural decoders use involves training a neural network to correct failures of a given decoder~\cite{chamberland_deep_2018,Baireuther2018machinelearning,Baireuther_2019,PhysRevA.99.052351}.
Note that when the decoder $ \add{D} $ produces a recovery operator $ \mathcal{R}_{\add{D}}(\vec{s}) $ on error $ \mathcal{E} $ corresponding to the error syndrome $ \vec{s} $,
applying the recovery operation on the error is equivalent to applying a logical operator $ \mathcal{L}_{\add{D}}(\vec{s}) \in \add{\set{\overline{I}, \overline{X}, \overline{Y}, \overline{Z}}} $ to the stabilizers $ \mathcal{G}_{\add{D}}(\vec{s}) $ of the original code state~\cite{PhysRevA.74.052333,PhysRevLett.104.050504},
\add{where the bar on top of the usual Pauli $I/X/Y/Z$ denotes the corresponding logical operator on the codespace}. This is summarised by
\begin{equation}
\label{eqn2}
\mathcal{R}_{\add{D}}(\vec{s}) \mathcal{E} = \mathcal{L}_{\add{D}}(\vec{s}) \mathcal{G}_{\add{D}}(\vec{s}).
\end{equation}
\add{The decoder $ D $} is successful if $ \mathcal{L}_{\add{D}}(\vec{s}) $ is the logical identity operator.
Thus, the goal of a neural decoder is to determine the logical Pauli operator applied to the code as a result of decoding and correct for it, if needed.

Our framework is similar in that it trains a neural network to learn the error syndrome that a given decoder fails on.
However, it only needs to learn which of the given decoders does not fail on that error syndrome rather than learning how a given decoder fails (i.e., which logical error the decoding operation results in).
Due to this difference, neural decoders trained to correct distinct decoders can be combined using our ensemble framework to boost error-correcting capability.

\add{
  We convey the intuition that differentiates the two techniques with the following example.
  Suppose that for a Monte Carlo simulation of the surface code under some noise model, there exists an error syndrome $ \vec{s} $ which, when decoded by $ D $ results in a logical error with the probability distribution,
\[
    \Pr[\mathcal{L}_{D}(\vec{s}) = \overline{P}] =
  \begin{cases}
    0.4, & \overline{P} = \overline{I} \\
    0.2, & \overline{P} \in \set{\overline{X}, \overline{Y}, \overline{Z}}
  \end{cases}
\]

  From the distribution above, an ideal neural decoder trained over $ D $ will learn that decoding is successful since $ \mathcal{L}_{D}(\vec{s}) $ is more likely to be $ \overline{I} $ than the other logical operators. Hence, it will not apply any correction over $ D $.
  However, an ideal neural ensemble can learn that another decoder $ D' $ with
  $ \Pr[\mathcal{L}_{D'}(\vec{s}) = \overline{I}] > \Pr[\mathcal{L}_{D}(\vec{s}) = \overline{I}] $
  will be more likely to succeed than $ D $. This would make it choose decoder $ D' $ for this syndrome.
}

\section{Results}
\label{sct:results}

To demonstrate the power of our techniques, we create an ensemble decoder consisting
of an MWPM decoder and an HDRG decoder, for surface codes of distance 5, 7, 9 and 11, assuming a depolarizing error model.
We generate a test data of 1 million samples using a Monte Carlo simulation
for physical error rates from $ 4\% $ to $ 16\% $.
We test our trained ensemble decoder on this data and compare its logical error rate with those of the MWPM and HDRG decoders individually.

\newcommand{\labelresult}[1]{\label{fig:L#1}}

\newcommand{\plotlogicalerror}[1]{
\begin{figure}
  \centering
  \includegraphics{#1}
  \caption{Logical error rates \add{[dimensionless]} for HDRG (triangle) MWPM (circle), and Neural ensemble (square) for distance \distance{} surface code. The pseudo-threshold for each decoder is indicated by the intersection with the \( y = x \) line (dashed).}
  \labelresult{\distance}
\end{figure}
}

\newcommand{\plotlogicalratio}[1]{
\begin{figure}
  \centering
  \includegraphics{#1}
  \caption{Ratio of logical error rates \add{[dimensionless]} of neural ensemble over MWPM for distances 5 (square), 7 (triangle), 9 (circle) and 11 (diamond).}
  \label{fig:ratio}
\end{figure}
}

\newcommand{\plotlogicalimprov}[1]{
\begin{figure}
  \centering
  \includegraphics{#1}
  \caption{Improvement in logical error rates \add{[dimensionless]} for neural ensemble over MWPM for distances 5 (square), 7 (triangle), 9 (circle) and 11 (diamond).}
  \label{fig:improv}
\end{figure}
}

\renewcommand{\distance}{5}
\renewcommand{\distanceroman}{V}

\plotlogicalerror{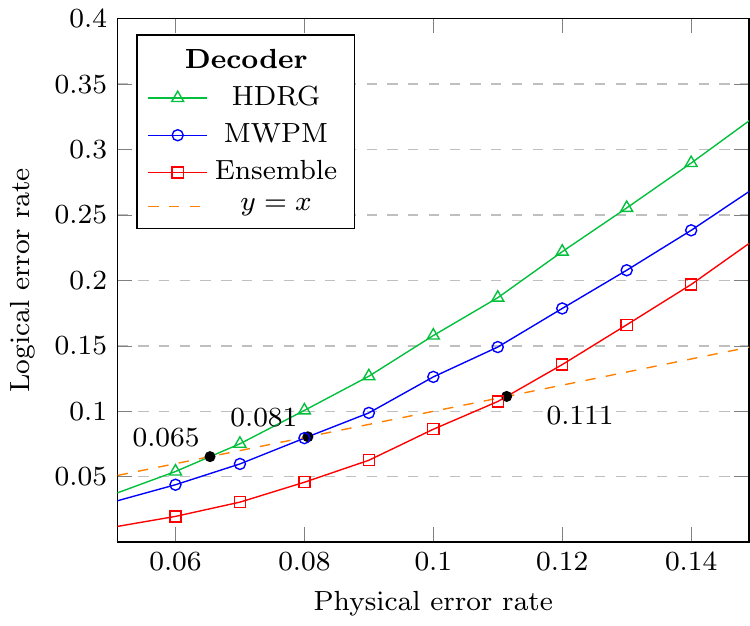}

\renewcommand{\distance}{7}
\renewcommand{\distanceroman}{VII}

\plotlogicalerror{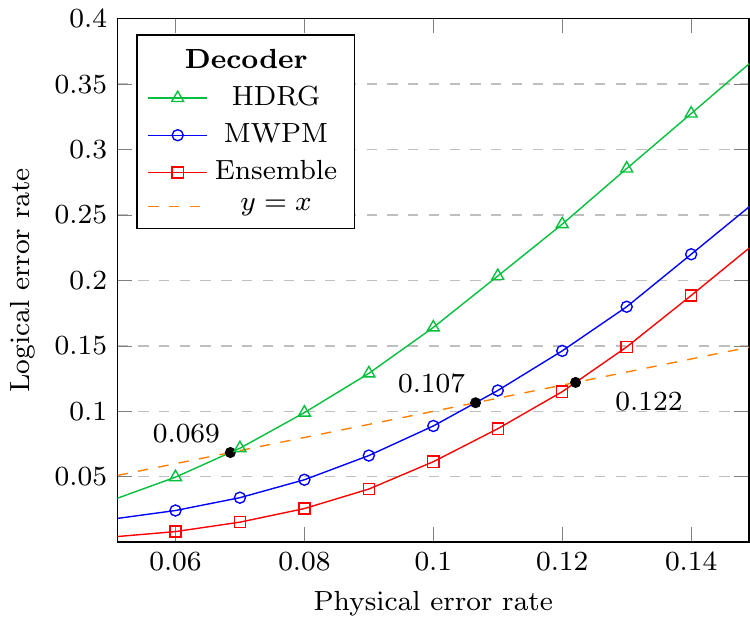}

\renewcommand{\distance}{9}
\renewcommand{\distanceroman}{IX}

\plotlogicalerror{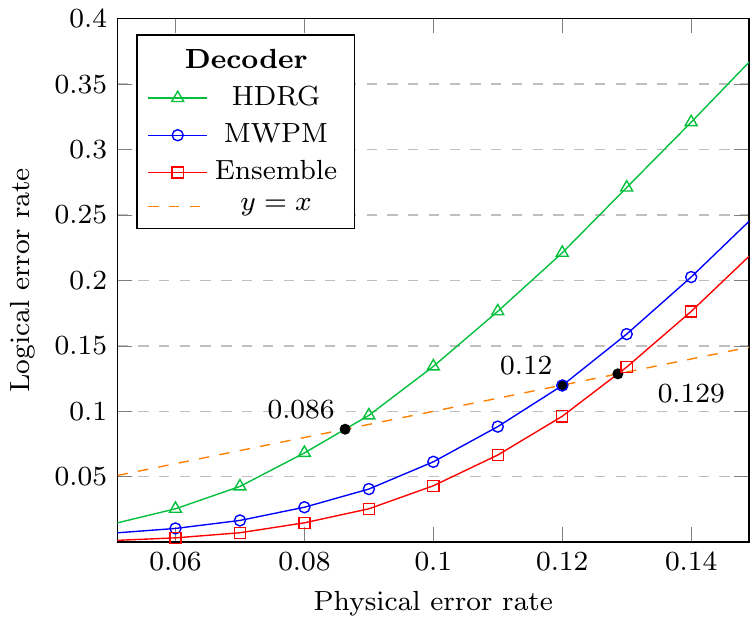}

\renewcommand{\distance}{11}
\renewcommand{\distanceroman}{XI}

\plotlogicalerror{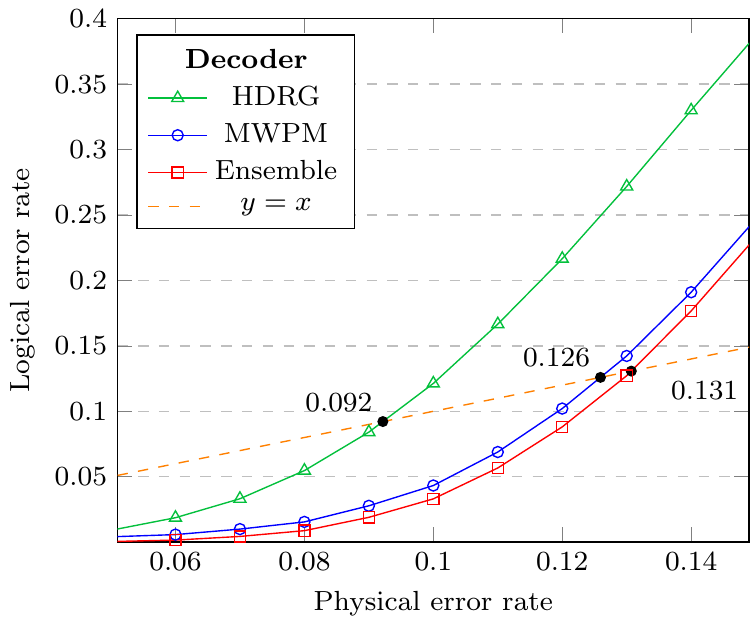}

\plotlogicalratio{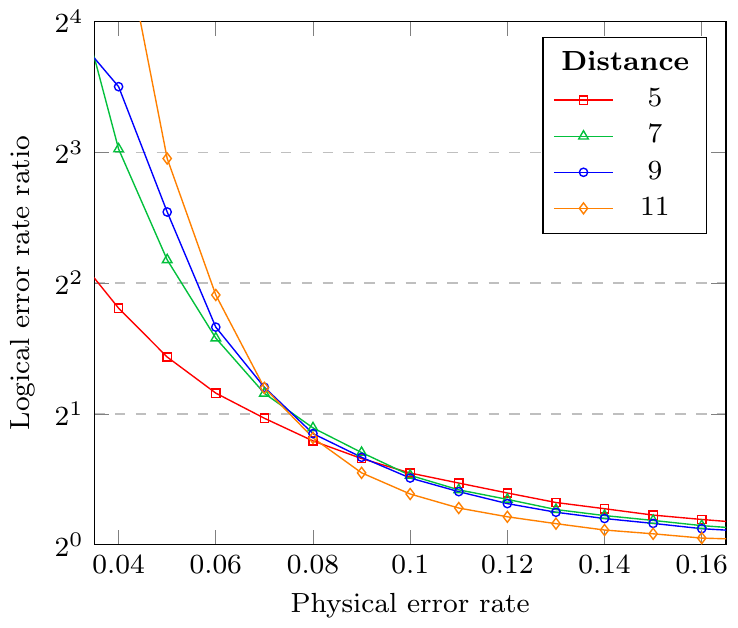}

\plotlogicalimprov{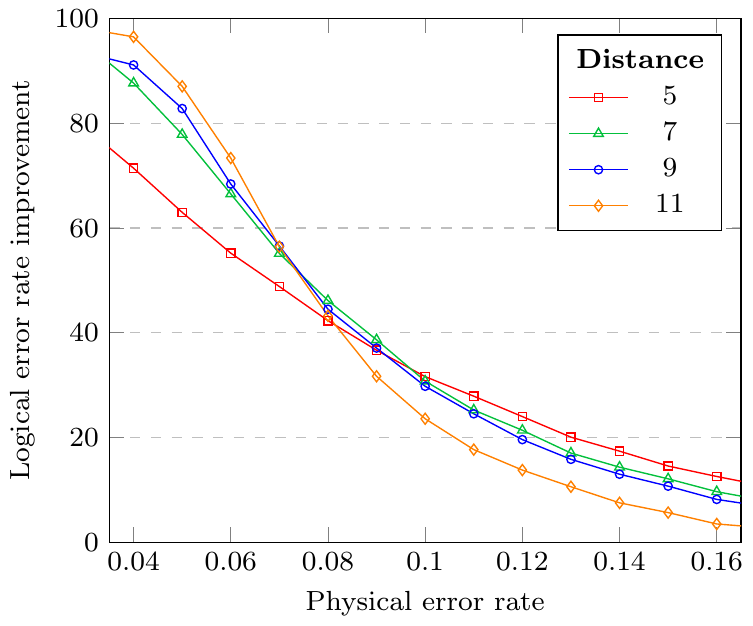}

We plot the logical error rates against the physical error rates in~\cref{fig:L5,fig:L7,fig:L9,fig:L11}.
We observe that the MWPM decoder correctly decodes more error syndromes than the HDRG decoder.
However, when the two decoders are combined using our ensemble framework, the HDRG
decoder boosts the MWPM success rates by a significant percentage.
A dotted $ y = x $ line intersects each decoder indicating the maximum error probability, namely the pseudo-threshold,
below which the logical error rate is lower than the physical error rate.
The results demonstrate an improved pseudo-threshold for the surface code when using an ensemble of the MWPM and HDRG decoders.
For example,~\cref{fig:L5} shows an improvement of \printpercent{\ensembleV}{\mwpmV}
in the pseudo-threshold for the distance 5 surface code.
The pseudo-threshold improvements are summarized in~\cref{tab:thres}.

\begin{table}
  \centering
  \begin{tabular}{ |c|c|c|c|c| }
    \hline
    Distance & HDRG  & MWPM  & Ensemble & Gain \\
    \hline
    5  & \printthreshold{\hdrgV}\%   & \printthreshold{\mwpmV}\%   & \printthreshold{\ensembleV}\%   & \printpercent{\ensembleV}{\mwpmV}     \\
    7  & \printthreshold{\hdrgVII}\% & \printthreshold{\mwpmVII}\% & \printthreshold{\ensembleVII}\% & \printpercent{\ensembleVII}{\mwpmVII} \\
    9  & \printthreshold{\hdrgIX}\%  & \printthreshold{\mwpmIX}\%  & \printthreshold{\ensembleIX}\%  & \printpercent{\ensembleIX}{\mwpmIX}   \\
    11 & \printthreshold{\hdrgXI}\%  & \printthreshold{\mwpmXI}\%  & \printthreshold{\ensembleXI}\%  & \printpercent{\ensembleXI}{\mwpmXI}   \\
    \hline
  \end{tabular}
  \caption{Pseudo-threshold comparison of HDRG, MWPM and our Neural Ensemble for various code distances.}
  \label{tab:thres}
\end{table}

Our results, see e.g. \cref{fig:L11}, also illustrate the difficulty of
retaining improvements in pseudo-threshold as the code distance increases.
Interestingly, our neural ensemble framework still provides a significant improvement over the
individual decoders when the physical error rate is lower than the pseudo-threshold.
To exemplify this fact, we plot the ratio of the logical error rates of the ensemble
over that of the MWPM decoder in~\cref{fig:ratio}, along with the improvement in
logical error rate for the ensemble over MWPM in~\cref{fig:improv}.
Note that as the physical error rate decreases, the ratio increases exponentially, which indicates that an ensemble is still a superior choice over the individual decoder for larger code distances
when utilized in a low-error environment, e.g. decoding the final layer of a concatenated code.

Our neural ensemble framework also benefits, similar to neural decoders~\cite{chamberland_deep_2018}, from the lack of assumption
made on the error distribution, which allows it to be applied to any underlying hardware and adapt to its corresponding
error distribution. In addition, ensemble decoding incurs minimal performance cost as the individual decoders
can be run in parallel until the ensemble outputs the decoder label to be used for decoding.
Evaluating an input on a trained neural network, known as inference, can be made very efficient using various techniques~\cite{chamberland_deep_2018, 8880492}.

We finish this section by mentioning the limitations of our approach.
Our framework suffers from some of the common problems faced by neural decoders.
In particular, our ensemble cannot handle the complexity of larger distances, as exemplified in \cref{fig:L11}
by the marginal gains to the pseudo-threshold in the case of the distance 11 surface code.
Due to the increase in training data needed for larger distances and the number of trainable parameters,
the training process becomes quickly computationally unfeasible.

Another limitation mentioned in~\cref{sbsct:gen} is concerned with applying a convolution-based architecture
to topological codes apart from the surface code. It is not obvious whether such a technique will work, as it requires a representation that efficiently encodes correlations in the error syndrome for the given topological code.

\section{Conclusion and open questions}
\label{sct:conclusion}

We have presented a simple and powerful framework for combining arbitrary decoders to boost decoding capability.
Our neural ensemble framework uses standard machine learning methods
for the classification of decoders with respect to error syndromes.
Empirical results listed in~\cref{tab:thres} demonstrate that an ensemble can improve the pseudo-thresholds by more than
\printpercent[0]{\ensembleV}{\mwpmV} for small distance surface codes compared to the MWPM decoder in the depolarizing noise model.
While our preliminary findings show potential of our techniques being translated into other
error models such as the circuit error model, a more thorough investigation is required.
The use of neural networks should allow our framework to adapt to the underlying error model
and thus be applicable to various underlying hardware.

One of the main constraints on decoding is the low coherence time of physical qubits.
This requires the use of very fast decoders which comes at the cost of error-correcting capability.

Our ensemble framework can be applied to two or more of such decoders to boost their capability
while adding minimal performance cost as noted in the previous section.
We have also discussed the difficulty of applying deep learning methods to large distances.
Despite this limitation, we observed that a hybrid approach using an ensemble during the outer layers of error-correction for a concatenated code, as an instance of a low error rate constraint, should prove to be fruitful.

Based on our experiments, we sketch possible avenues for future research.
One direction to explore is finding a method for constructing the set of decoders that maximize the error-correcting capability of the corresponding ensemble decoder built out of them.
One interesting example constitutes of combining efficient decoders specializing in different error models to create a general purpose decoder with the ability to handle non-idealized noise processes.

More sophisticated machine learning techniques may
also improve the classification task.
In particular, noise models allowing measurement errors involve a time-valued correlation that benefit
from recurrent neural network techniques~\cite{Baireuther2018machinelearning,chamberland_deep_2018}.
Our framework can be extended to use recurrent CNNs to account for measurement errors.
\add{We also mentioned in \cref{sbsct:train}, that clustering the error syndromes (or the corresponding Pauli noise), under an appropriate distance metric, could help with the classification.}

\add{
  As briefly discussed in \cref{sbsct:compare}, combining ideas from neural decoders with our neural ensemble framework could significantly improve its capability. This could entail creating an ensemble over neural decoders or adapting their architectural techniques.
}

\add{
  It would also be interesting to identify patterns in error syndromes that make one decoder more likely to successfully decode the syndrome over another. This might allow us to implement an efficient pattern detection algorithm to cover most syndromes where the decoders differ, and use it instead of the neural ensemble to classify the decoders to avoid the expensive floating point operations of evaluating neural networks. And while it may be hard to identify such patterns for one pair of decoders, it might be easier for another pair.
}

Recently, various methods to go beyond small distances for neural decoders have been proposed~\cite{varsamopoulos2019decoding, ni2018neural}. Those might be applicable to our neural network architecture and their investigation is the subject of future research.

Finally, another direction one may explore is applying the ensemble framework
for other types of codes (e.g. color codes, low-density parity-check (LDPC) codes etc.).
As discussed in~\cref{sbsct:gen}, this requires a good representation
of the error syndrome that encodes the correlations between qubits.

We hope the ensemble framework is a valuable tool for obtaining fast near-optimal decoders for topological codes.

\begin{acknowledgments}
We thank Pooya Ronagh for useful discussions regarding methods for improving the training of the machine learning model. \add{We also thank the anonymous referee for her/his very useful suggestions that improved the quality of this manuscript.}
M.S. acknowledges support from the Undergraduate Research Award at IQC. S.Z. acknowledges support from ``NSERC CREATE in Building a Workforce for the Cryptographic Infrastructure of the 21st Century (CryptoWorks21)'' and ``Quantum Information Science, Cryptography and Privacy''  NSERC Discovery Grants Program.
We thank IQC for providing computational resources. We acknowledge support from NSERC and CIFAR. IQC is supported in part by the Government of Canada and the Province of Ontario.
\end{acknowledgments}

\bibliographystyle{apsrev4-1}
%

\end{document}